\documentclass[aps,twocolumn,showpacs,preprintnumbers,amssymb]{revtex4}

\newcommand{\be}{\begin{eqnarray}}
\newcommand{\ee}{\end{eqnarray}}

\newcommand{\eins}{\mbox{$1 \hspace{-1.0mm}  {\bf l}$}}
\def\bea{\begin{eqnarray}}
\def\eea{\end{eqnarray}}
\def\C{\hbox{$\mit I$\kern-.7em$\mit C$}}
\def\N{\hbox{$\mit I$\kern-.3em$\mit N$}}

\def\tr{{\rm tr}}

\usepackage{dcolumn} 
\usepackage{bm} 

\usepackage{epsfig}

\begin{document}

\title{Multiparticle entanglement purification for graph states}

\author{W. D\"{u}r, H. Aschauer and H.-J. Briegel}

\affiliation{
Sektion Physik, Ludwig-Maximilians-Universit\"at M\"unchen, Theresienstr.\ 37, D-80333 M\"unchen, Germany.}

\date{\today}

\begin{abstract}
We introduce a class of multiparticle entanglement purification protocols that allow us to distill a large class of entangled states. These include cluster states, GHZ states and various error correction codes all of which belong to the class of two-colorable graph states. We analyze these schemes under realistic conditions and observe that they are scalable, i.e. the threshold value for imperfect local operations does not depend on the number of parties for many of these states. 
When compared to schemes based on bipartite entanglement purification, the protocol is more efficient and the achievable quality of the purified states is larger. 
As an application we discuss an experimental realization of the protocol in optical lattices which allows one to purify cluster states.
\end{abstract}


\maketitle

Entangled states of multi-partite systems are expected to play a significant role for applications in quantum computation and quantum communication. Quantum error-correcting codes, as the most prominent example, are highly entangled states which are used to encode quantum information and to protect it against the damaging effects of decoherence.  It has been shown that similar entangled states may be used in  multi-party communication scenarios beyond teleportation and quantum key distribution \cite{Be93}, such as secret sharing or secure function evaluation \cite{secure,Cr02}, but also in more practical applications such as the improvement of existing frequency standards \cite{Wi93}. In the context of quantum computation, specific multi-partite entangled states --- the so-called cluster states \cite{Ra01a} - have even been shown to constitute a  universal resource for measurement-based quantum computation \cite{Ra01b}. These states can be created e.g.\ via an Ising interaction between neighboring particles on a lattice. A specific realization of such a system is based on neutral atoms in an optical lattice where, starting from a Mott-insulator state \cite{Bl01}, cluster states could be created by a simple interferometric process \cite{Ra01a,Bl02}.

When talking about entanglement, it is mandatory to consider the effect of decoherence.  Finite temperature (in the case of  interacting particles)  or a noisy communication channel (in the case of distributed quantum systems) lead to decoherence in the state space of the information carriers, with a rate that increases typically in proportion with the number of particles. Therefore only mixed states rather than pure states will be available, and the achievable fidelity is expected to decrease exponentially both with time and with the size of the system. It is thus not clear whether an entangled state of a large number of particles can be created and maintained in practice.

Standard methods have been developed to stabilize quantum states against the detrimental effects of decoherence and noise. Quantum error correction is one general method, but the acceptable noise level for the quantum operations is extremely small (of the order of $10^{-4}$) \cite{Pr97}. Entanglement purification \cite{Be96,De96} is an alternative, but protocols exist only for the purification of a special type of states, namely states which are equivalent, up to local unitary operations, to the so-called ``cat'' states  $|0\rangle^{\otimes N} + |1\rangle^{\otimes N}$ \cite{Be96,De96,Mu98,Sm00}. 
Furthermore, it needs to be investigated whether these protocols are at all applicable under realistic conditions. While in the bipartite case certain entanglement purification protocols turn out to be remarkably robust against the influence of imperfect local operations -- allowing for threshold values at the order of several percent \cite{Br98} --, it is not clear whether this is also true in the multipartite case. In particular, it is not known how the threshold for the required precision of local control operations scales with the size of the system, a crucial question if one wants to establish and maintain quantum entanglement in larger systems.

In this paper, we introduce a class of entanglement purification protocols, which can be used to purify a large class of  multi-partite entangled states, in particular all two--colorable graph states. 
These states include, for example, various quantum error correcting codes \cite{Sc01}, the generalized GHZ states (or cat states), and the cluster states, all of which are resources for the applications mentioned above. 
We analyze these protocols under realistic conditions, i.e.\ when the local operations required in the purification process are imperfect.
We find that: 
{\em (i)} The value of the purification threshold, i.e.\ the acceptable noise level for local operations under which the protocol still purifies, does not depend on the number of particles, but is influenced by the maximal degree of the graph associated with the state.
{\em (ii)} The described  multi-particle entanglement purification protocols (MEPP) are generally not only more efficient than any method based on bipartite entanglement purification protocols (BEPP), but in the case of imperfect local operations the achievable fidelity is even higher. 
{\em (iii)} The entanglement of the purified multipartite states is private \cite{As99}, thus providing a secure resource for applications in multiparty            communication scenarios \cite{Cr02}. 
We close by describing how MEPP could be implemented experimentally in optical lattices, allowing one to increase the fidelity of 
cluster states.


We start out by introducing two--colorable graph states, and purification protocols which allow one to purify all states of this kind. 
Consider a graph $G=(V,E)$ which is a set of vertices $V$ connected in a specific way by edges $E$ specifying the neighborhood relation between vertices. With this graph we associate $N=|V|$ commuting correlation operators
$K_j= \sigma_x^{(j)} \prod_{\{k,j\} \in E} \sigma_z^{(k)}.$
The joint eigenstates of $K_j$, $|\Psi_{\mu_1\mu_2\ldots \mu_N}\rangle$, $\mu_j \in \{0,1\}$, the graph states associated with $G$, fulfill the eigenvalue equation $K_j |\Psi_{\mu_1\mu_2\ldots \mu_N}\rangle = (-1)^{\mu_j}|\Psi_{\mu_1\mu_2\ldots \mu_N}\rangle ~\forall j$. Note that $\{|\Psi_{\mu_1\mu_2\ldots \mu_N}\rangle\}_G$ form a basis in ${\cal H}=(\C^2)^{\otimes N}$. 
We restrict our attention to two--colorable graphs, that are graphs for which a partition of the vertices into two disjoint sets $V_A \cup V_B=V$ exists such that no vertices within a set are connected by edges. 
The states arising from such two--colorable graphs, which we call {\em two--colorable graph states} (TCGS), include various interesting MES, ranging from $N$-particle GHZ states over codewords for quantum error correction codes to cluster states. In what follows, we will introduce protocols which allow one to purify arbitrary TCGS. That is, given $M$ copies of an arbitrary $N$-partite mixed state $\rho$, we establish for each two--colorable graph $G$ a local protocol which is capable of creating the pure state $|\Psi_{\vec 0}\rangle_G$ as output state, provided the initial state $\rho$ fulfills certain requirements (e.g.\ has sufficiently high fidelity with respect to $|\Psi_{\vec 0}\rangle_G$).


Let us consider an arbitrary but fixed two--colorable graph $G$ with vertices $V=V_A\cup V_B$, $N_A\equiv|V_A|, N_B\equiv|V_B|$ and $N=N_A+N_B$ spatially distinct parties each holding one of the $N$ particles belonging to a general mixed state $\rho$. One can depolarize the state $\rho$ to a state $\rho_G$ which is diagonal in the graph-state basis $\{|\Psi_{\mu_1\mu_2\ldots \mu_N}\rangle\}_G$ without changing the diagonal elements by a  
probabilistically applying the local operations corresponding to the operators $K_j$ \cite{As03}.
That is, without loss of generality we can consider mixed states diagonal in the graph-state basis, 
\be
\rho_G=\sum_{\vec\mu_A,\vec\mu_B} \lambda_{\vec\mu_A,\vec\mu_B} |\Psi_{\vec\mu_A,\vec\mu_B}\rangle\langle \Psi_{\vec\mu_A,\vec\mu_B}|,
\ee
where we introduced the multi indices $\vec\mu_A, \vec\mu_B$ with e.g.\ ${\vec\mu_A} \equiv \mu_{i_1}\mu_{i_2} \ldots \mu_{i_{N_A}}$ indicating all vertices in the set $V_A$. We now introduce two protocols $P1, P2$ acting on two identical copies $\rho_1=\rho_2=\rho_G$, $\rho_{12}\equiv\rho_1\otimes \rho_2$. In protocol $P1$, all parties which belong to set $V_A$ [$V_B$] apply local CNOT operations \cite{noteCNOT} to their particles, with the particle belonging to $\rho_1$ [$\rho_2$] as source, $\rho_2$ [$\rho_1$] as target. 
From the eigenvalue equations for $|\Psi_{\vec\mu_A,\vec\mu_B}\rangle$ follows that the action of the local CNOT operations is given by  
\be
|\Psi_{\vec \mu_A,\vec\mu_B}\rangle|\Psi_{\vec \nu_A,\vec\nu_B}\rangle\rightarrow|\Psi_{\vec\mu_A,\vec\mu_B\oplus\vec\nu_B}\rangle|\Psi_{\vec\nu_A\oplus \vec\mu_A,\vec\nu_B}\rangle,
\label{psitopsi}
\ee
where $\vec\mu_A\oplus\vec\nu_A$ denotes bitwise addition modulo 2. For instance, if $\vec\mu_A=\mu_1\mu_3\mu_5$, $\vec\mu_A\oplus\vec\nu_A=\mu_1\oplus\nu_1,\mu_3\oplus\nu_3,\mu_5\oplus\nu_5$. 
All particles of $\rho_2$ belonging to set $V_A$ [$V_B$] are then measured in the eigenbasis of $\sigma_x$ [$\sigma_z$] respectively, yielding results $(-1)^{\xi_j}$ [$(-1)^{\zeta_k}$] with $\xi_j,\zeta_k \in\{0,1\}$. The first state is only kept if the measurement outcomes fulfill $(\xi_j+\sum_{\{k,j\}\in E}\zeta_k){\rm mod}2=0 ~\forall j$ which implies $\vec\mu_A\oplus\vec\nu_A=\vec 0$. Eq.~(\ref{psitopsi}) makes it evident that in this procedure information about the first state $\rho_1$  -- encoded into $\vec \mu_A,\vec\mu_B$ -- is transferred to the second state $\rho_2$ and revealed by the measurement. This is in analogy to the standard recurrence protocols \cite{Be96,De96,Mu98,Sm00} and it is the key point to purifying mixed states.

The resulting state $\tilde\rho$ after this procedure is again diagonal in the graph-state basis, with new coefficients
$\tilde\lambda_{{\vec \gamma_A},{\vec \gamma_B}} =\sum_{\{({\vec \nu}_B, {\vec \mu}_B) | {\vec \nu}_B \oplus{\vec \mu}_B={\vec \gamma_B}\}} \frac{1}{2K}\lambda_{{\vec \gamma_A},{\vec \nu_B}}\lambda_{{\vec \gamma_A},{\vec \mu_B}},$
where $K$ is a normalization constant such that $\tr(\tilde \rho)=1$. In the protocol $P2$, the role of sets $V_A$ and $V_B$ is interchanged, which leads to $\tilde\lambda'_{{\vec \gamma_A},{\vec \gamma_B}} =\sum_{\{({\vec \nu}_A, {\vec \mu}_A)| {\vec \nu}_A \oplus{\vec \mu}_A={\vec \gamma_A}\}} \frac{1}{2K}\lambda_{{\vec \nu_A},{\vec \gamma_B}}\lambda_{{\vec \mu_A},{\vec \gamma_B}}$. The total purification protocol corresponds to an iterative application of subprotocols $P1$ and $P2$, always using two identical copies of the multiparticle states obtained in the previous round as input states. 
For GHZ states, this protocol is equivalent to the one introduced in Ref. \cite{Mu98}, and further analyzed in Ref. \cite{Sm00}, and may thus be viewed as a generalization to the class of all TCGS. 


To gain analytical insight into this procedure, we consider the example of rank $2^{N_A}$ mixed states of the form $\rho_{\cal A}\equiv \sum_{\vec\mu_A} \lambda_{\vec\mu_A,\vec 0} |\Psi_{\vec \mu_A,\vec 0}\rangle \langle \Psi_{\vec \mu_A,\vec 0}|$. The application of protocol $P1$ leads to another state of this form with new coefficients $\tilde\lambda_{\vec\mu_A,\vec 0} = \lambda_{\vec\mu_A,\vec 0}^2/K$ and $K=\sum_{\vec\mu_A} \lambda_{\vec\mu_A,\vec 0}^2$. That is, the largest coefficient is amplified with respect to the other ones and iteration of the protocol allows one to produce pure graph states. Consider for instance the one parameter family $\rho_{\cal A}(F)$ with $\lambda_{\vec 0,\vec 0}=F$, $\lambda_{\vec\mu_A,\vec 0}=(1-F)/(2^{N_A}-1)$ for $\vec \mu_A \not=\vec 0$, where $F$ is the fidelity of the desired state. Application of $P1$ preserves the structure of those states and leads to $\tilde F = F^2/[F^2+ (1-F)^2/(2^{N_A}-1)]$. This map 
 has 
$\tilde F=1$ as attracting fixed point for $F\geq 1/2^{N_A}$. In these examples, application of protocol $P1$ alone is sufficient as only information about $\vec \mu_A$ has to be extracted. 


For general full rank mixed states, both $P1$ and $P2$ will have to be applied in order to reveal information about $\vec \mu_A$ [$\vec \mu_B$] respectively. While $P1$ increases the weight of coefficients $\lambda_{\vec 0,\vec\mu_B}$, $P2$ amplifies coefficients $\lambda_{\vec\mu_A,\vec 0}$ which together leads to the amplification of $\lambda_{\vec 0,\vec 0}$ given the initial fidelity is sufficiently high. The action of $P1$, $P2$ correspond to non-linear mappings of a large number of independent variables (in total $2^N-1$), which makes an analytic treatment of the purification protocol very difficult. We have however analyzed the protocol numerically and found that it is indeed capable to purify noisy TCGS arising from various kinds of noise. In the following, we will concentrate on GHZ states and 1D cluster states. The graph corresponding to the $N$-particle GHZ state is given by $V_A=\{1\},V_B=\{2,3,\ldots,N\}$ with edges $\{1,k\}, k\in V_B$, while 1D cluster states correspond to a graph where vertex $k$ is connected to vertex $k+1, \forall k$ 
and sets $V_A$ [$V_B$] are formed by all odd [even] vertices respectively. 
Our results are however not restricted to these specific graphs.

\begin{figure}[ht]
\begin{picture}(230,90)
\put(-5,-5){\epsfxsize=230pt\epsffile[-14 155 1090 601]{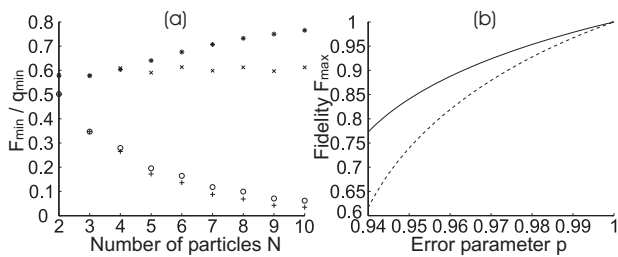}}
\end{picture}
\caption[]{(a): Minimal value of fidelity $F_{\rm min}$ $+$ [$\circ$] and parameter $q_{\rm min}$ $\times$ [$*$] for linear cluster states [GHZ states] for different number of particles $N$ and perfect local operations. (b): Achievable fidelity of a linear cluster state with $N=4$ using direct MEPP (solid line) and conservative upper bound for methods based on BEPP (dashed line) for different errors in local operations $p$.}
\label{Fig_Fminqmin}
\label{Fig_comparebipartite}
\end{figure}

We consider noisy TCGS arising naturally in a multipartite scenario where each of the $N$ particles constituting $|\Psi_{\vec 0}\rangle$ is subject to decoherence, arising e.g.\ from sending the particles through noisy quantum channels. We consider depolarizing channels with noise parameter $q$ described by
\be
{\cal E}_k\rho= q \rho +(1-q)/2\eins_k \otimes \tr_k(\rho), \label{whitenoise}
\ee
where in this case the channel is acting on particle $k$. The resulting multipartite state is of the form $\rho(q) = {\cal E}_1{\cal E}_2 \ldots {\cal E}_N |\Psi_{\vec 0}\rangle\langle \Psi_{\vec 0}|$ which we take as input states for our MEPP. It turns out that the tolerable amount of white noise per particle 
such that our MEPP can still be successfully applied is for linear cluster states essentially independent of the number $N$ of particles, while for GHZ states this value decreases exponentially with $N$ (see Fig.~\ref{Fig_Fminqmin}a). This implies that, for the GHZ state, the required quality of the channel over which particles are distributed, such that the resulting state is still purifiable, increases with $N$. Note that this different behavior is not reflected in the minimum fidelity $F_{\rm min}$ which in both cases decreases exponentially (Fig.~\ref{Fig_Fminqmin}a). We remark that for all states $\rho(q)$ where $q$ is larger than the threshold value, the protocol successfully converges towards the desired state $|\Psi_{\vec 0}\rangle$. For more general states, conditions for convergence are difficult to determine due to the large number of parameters, even though convergence itself can straightforwardly be checked numerically.
 Also for noisy states of the form $\rho(x)=x|\Psi_{\vec 0}\rangle\langle \Psi_{\vec 0}|+(1-x) /2^N\eins$, i.e.\ mixtures of the desired state with a                   completely depolarized state, the situation is similar, i.e.\ $F_{\rm min}$ decreases exponentially with $N$.

Until now we have assumed that local operations -- in particular CNOT operations -- are perfect. In practice, however, these operations as well as measurements will be imperfect. We now investigate the influence of errors in the local operations on the MEPP.  We will consider a simple error model where local two-qubit operations are described by the completely positive map
${\cal E}_{U_{jk}}\rho = U_{jk} [{\cal E}_j {\cal E}_k\rho] U_{jk}^{\dagger},$
where ${\cal E}_k, {\cal E}_j$ are given by Eq.~(\ref{whitenoise}) with error parameter $p$. That is, an imperfect operation is described by first applying local white noise with probability $(1-p)$ independently on the qubits, followed by the perfect unitary operation. 
We have also investigated more general error models, e.g.\ 
correlated white noise, and errors in the measurement process, observing essentially the same behavior as for this simple model. 
 
\begin{figure}[ht]
\begin{picture}(230,90)
\put(-5,-5){\epsfxsize=230pt\epsffile[-6 155 1110 601 ]{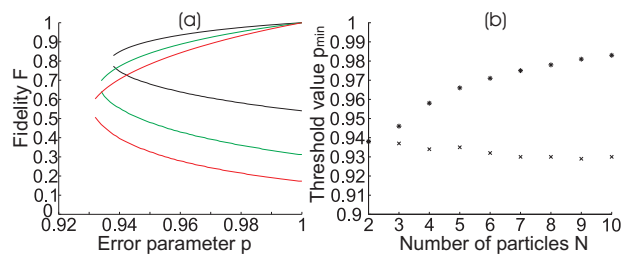}}
\end{picture}
\caption[]{(a): Maximal reachable fidelity $F_{\rm max}$ and minimal required fidelity $F_{\rm min}$ plotted against error parameter $p$ (local operations) for density operators arising from single-qubit white noise. Curves from top to bottom correspond to linear cluster states with $N=2,4,6$ particles. (b): Threshold value for errors in local operations $p_{\rm min}$ for GHZ states ($*$) and linear cluster states ($\times$) with different number of particles $N$.} 
\label{Fig_FminFmax}
\label{Fig_thresholdp}
\end{figure}

We have investigated the dependence of the minimal required fidelity and the maximal reachable fidelity for linear cluster states of different length on the error parameter $p$ (see Fig.~\ref{Fig_FminFmax}a) as well as the threshold value $p_{\rm min}$ until which our MEPP can be successfully applied (Fig~\ref{Fig_thresholdp}b). As can be seen from Fig.~\ref{Fig_thresholdp}b, $p_{\rm min}$ is almost independent of the number of particles $N$, it even seems that for larger $N$ the tolerable amount of noise per operation is larger. For GHZ states, in contrast, the threshold value $p_{\rm min}$ increases with $N$, i.e.\ the requirements to purify GHZ states with larger $N$ are more stringent. The qualitatively different behavior of $p_{\rm min}$ for these two TCGS can be understood within a restricted error model. We consider noisy operations where only particles in $V_B$ are subjected to bit-flip errors -- described by the map $\tilde{\cal E}_k\rho = p\rho+(1-p)/2(\rho+\sigma_x\rho\sigma_x)$ -- and entanglement purification is performed by applying protocol $P1$. 
For GHZ states we have that $\sigma_x^{(k_B)}$ is equivalent to phase flip errors in particle 1, $\sigma_z^{A_1}$ $\forall k_B\in V_B$ as can be seen from the eigenvalue equations, while for closed linear cluster states $\sigma_x^{(2k)}$ corresponds to $\sigma_z^{(2k-1)}\sigma_z^{(2k+1)}$. Thus for GHZ states, errors at $N-1$ particles affect particle 1 and the errors accumulate. For linear cluster states, however, each particle in $V_A$ is only affected by errors at two other particles in $V_B$, independent of $N$. One can show analytically that even for this restricted class of errors, for GHZ states $p_{\rm min}$ increases exponentially with $N$, $p_{\rm min}=1/2^{1/(N-1)}$. For linear cluster states one finds that for input states of the form $\rho_{\cal A}(F)$ and a fixed error rate, e.g.\ $p=0.8$, one application of the protocol $P1$ increase the fidelity $F\equiv x+(1-x)/2^{N/2}$ in the range $2^{-0.33 N}\leq x \leq 2^{-0.009 N}$ \cite{As03}. That is, for each $N$ there exists a finite regime where entanglement purification is possible and the threshold value $p_{\rm min}$ converges for large $N$ towards $p_{\rm min}\approx 0.4938$.

We have also performed numerical simulations for various other two--colorable graphs corresponding e.g.\ to 2D cluster states and simple CSS codes \cite{Sc01}, observing a similar behavior as for linear cluster states. Since graphs corresponding to a concatenation of such CSS codes are also two--colorable \cite{As03}, our approach may be used to purify entire encoding circuits used in fault tolerant quantum computation. While in the general case the structure of the graph -- in particular its maximum degree -- influences the dependence of the threshold value $p_{\rm min}$ of the corresponding purification protocol, the exact dependence of $p_{\rm min}$ on the graph is still an open problem.

Let us compare this MEPP with schemes based on bipartite entanglement purification. By this we mean a scheme which produces bipartite entangled states by means of BEPP, which are then used to create a MES by some means, e.g.\ by teleportation. Even for the best known BEPP with respect to maximal reachable fidelity under imperfect local operations -- which is the protocol introduced in Ref. \cite{De96} -- and under the conservative assumption that the local operations involved in the creation of multiparticle entangled states from (noisy) entangled pairs are itself error free, the maximal reachable fidelity $F_{\rm max}^{\rm MP}$ for our MEPP is considerably larger than the upper bound for methods based on BEPP, as can be seen in Fig.~\ref{Fig_comparebipartite}b. Note that the upper bound in the latter case solely depends on the fixed point of the BEPP. This shows that there are tasks that can be solved with multi-lateral purification (i.e. the efficiency of the process is non--zero) but not with bi-lateral purification (i.e. the efficiency is zero), where the efficiency is given by the ratio of surviving copies with the desired fidelity to the number of initial copies (see Ref. \cite{As03} for details).


Finally, we propose an experimental realization of MEPP with neutral atoms in optical lattices. We show that MEPP can be used in such systems to increase the fidelity of cluster states. In particular, we consider the purification of 1D cluster states in a 2D lattice.
Remarkably, even if the same imperfect operations are involved in the creation of the cluster state and in the purification process, MEPP allows one to enhance the achievable fidelity of the state \cite{As03}. For the experiment, consider a two-dimensional $N \times N$ optical lattice filled with one atom per lattice site. Internal states of the atoms -- which constitute the qubits -- can be manipulated by means of laser pulses. Interactions between neighboring atoms take place e.g.\ by state-selectively shifting the lattice, leading to a state dependent collisional phase arising from controlled cold collisions \cite{Ja98}. The interaction Hamiltonian describing a lattice shift in the $x$-direction is given by $H_x=g(t) \sum_{k,l} (1+\sigma_z^{(k,l)})/2 \otimes (1-\sigma_z^{(k+1,l)})/2$, where $(k,l)$ labels the $(x,y)$-coordinate of the atom and $g(t)$ is a time dependent coupling parameter. Note that for $\int g(t) dt=\pi$, such an interaction produces $N$ copies of one dimensional cluster states along the $x$-direction of the lattice when applied to states of the form $(|0\rangle+|1\rangle)^{\otimes N}$. These states can then be purified by using lattice shifts along the $y$-direction: By applying local laser pulses to the individual atoms, the resulting interaction for $\int g(t) dt=\pi$ can be converted into the proper CNOT operations between pairs of atoms, thereby realizing (\ref{psitopsi}). Using a total of two lattice shifts, this process can even be parallelized such that $N/2$ pairs of cluster states are purified simultaneously. The resulting states after a successful measurement can further be purified by applying protocol $P2$ in a similar way. 


In this paper we have introduced entanglement purification protocols which are capable of purifying a wide range of multiparticle states, namely all TCGS. In the case of noisy operations, these multiparticle entanglement purification protocols are not only remarkably robust, but they allow one to reach higher fidelities than known schemes based on bipartite purification. We also found that for cluster states the threshold value for imperfect local operations is independent of the number of particles. Our results are a step towards practical applications based on multiparticle entangled states.


We thank R. Raussendorf, M. Grassl and M. Hein for discussions. 
This work was supported by the DFG and the European Union (HPMF-CT-2001-01209 (W.D.), IST-2001-38877,-39227).







\begin{thebibliography}{99}
\bibitem{noteCNOT}
The CNOT operation is defined by $|i\rangle_A|j\rangle_B \rightarrow |i\rangle_A|i\oplus
j\rangle_B$, where $\oplus$ denotes addition modulo 2.
\end{thebibliography}
\end{document}